# Considerations for the Integration of Clinical Trials and Real-World Data

**Authors:**

Sky Qiu[1 2 3], Charles E. Barr[4], Lauren Dang[5], Larry Han[6 7], Kajsa Kvist[8], Hana Lee[9], Andrew Mertens[10 2], Nerissa Nance[8 2], Lei Nie[9], Kara Rudolph[11], Xu Shi[12], Jens Tarp[8], Salina P. Waddy[13], Kenneth Wiley[13], Andy Wilson[14 15], Margot Lisa Jing Yann[16], Zhiwei Zhang[17], Tianyue Zhou[1 2], Maya Petersen[1 2], and Mark van der Laan[1 2]

**Correspondence:**

Sky Qiu[1 2 3], sky.qiu@berkeley.edu

**Affiliations:**

[1] Division of Biostatistics, School of Public Health, University of California, Berkeley, California, United States

[2] Center for Targeted Machine Learning and Causal Inference, School of Public Health, University of California, Berkeley, California, United States

[3] Targeted ML Solutions, Inc., Cambridge, Massachusetts, United States

[4] Adaptic Health, Mountain View, California, United States

[5] National Institute of Allergy and Infectious Diseases, Bethesda, Maryland, United States

[6] Department of Public Health and Health Sciences, Northeastern University, Boston, Massachusetts, United States

[7] Department of Biostatistics, School of Public Health, Brown University, Providence, Rhode Island, United States

[8] Novo Nordisk A/S, Bagsvaerd, Denmark

[9] Center for Drug Evaluation and Research, U.S. Food and Drug Administration (FDA), Silver Spring, Maryland, United States

[10] Division of Epidemiology and Biostatistics, School of Public Health, University of California, Berkeley, California, United States

[11] Department of Epidemiology, Columbia University Mailman School of Public Health, New York City, New York, United States

[12] Department of Biostatistics, School of Public Health, University of Michigan, Ann Arbor, Michigan, United States

[13] National Institutes of Health, Bethesda, Maryland, United States

[14] University of Utah, Salt Lake City, Utah, United States

[15] Tao of RWD, Salt Lake City, Utah, United States

[16] Forum for Collaborative Research, School of Public Health, University of California, Berkeley, California, United States

[17] Gilead Sciences, Foster City, California, United States

**Disclosures:** NN is employed by Novo Nordisk A/S. The views are those of the author(s) and do not necessarily represent the official views of, nor an endorsement, by FDA, HHS, or the U.S. Government.

## Abstract

As clinical decision-making increasingly moves toward individualized and context-specific treatment recommendations, reliance on any single evidence source, randomized or observational, may be insufficient. Principled integration of randomized controlled trials and real-world data, grounded in explicit causal frameworks, offers a path toward evidence that is both internally credible and externally relevant. In this article, we describe distinct objectives for the integration of randomized controlled trials and real-world data and discuss how these objectives shape key design and analytic considerations, illustrating the resulting choices through example estimands. We highlight practical issues that commonly arise in applied settings, including data relevance and curation, cross-source comparability, estimand specification, and sensitivity analysis. We aim for this article to help readers evaluate and implement principled approaches to integrating randomized controlled trials and real-world data in ways that can support more reliable treatment recommendations while maintaining regulatory-grade evidentiary standards.

## 1 Introduction

Regulatory decision-making has traditionally relied heavily on randomized controlled trials (RCTs) as the gold standard for establishing treatment efficacy and safety[1]. However, as the landscape of drug development evolves, the limitations of considering evidence from RCT studies alone are increasingly recognized, prompting greater interest in methodologies that integrate real-world data (RWD) and RCT data in the evidence generation process[2–6]. The evidence generation ecosystem is evolving as well from meta-analysis tools that integrate findings at the study-level to statistical techniques that data-adaptively integrate patient-level data from different studies[7]. These methodological advances are enabling a paradigm shift in evidence generation and synthesis. Rather than asking decision makers (clinicians and regulators) to “consider the totality of evidence” and mentally integrate disparate trials, observational studies, and meta-analyses qualitatively, we can increasingly provide an explicit quantitative synthesis, together with the context and sensitivity analyses needed to interpret it.

Recent methodologic advance also makes it possible to account for potential biases arising from various misalignments between data sources, including distributional shifts and structural differences in how data are generated and recorded, such as differences in patient populations and eligibility criteria, treatment implementation and adherence, endpoint definitions, confounding mechanisms, and even coding systems[8,9]. The ultimate objective of these methods is to integrate evidence from disparate sources to best inform a target scientific question, which is especially relevant in settings where no single data source is fully sufficient on its own.

Study designs that integrate or augment RCTs with external RWD often capitalize on three key strengths of RWD: larger sample sizes, greater patient diversity, and longer follow-up times. Such integration may enhance statistical power and improve post-marketing safety analyses of rare outcomes. RWD often includes a broader and more diverse patient population, aiding in both generalizability to understudied groups and transportability of treatment effects to target populations beyond the trial-eligible population[10,11]. These integrated designs could facilitate earlier patient access to potentially safer and more effective drugs[12]. Linking RCT patients with their RWD generated post-trial may also allow researchers to study long-term treatment effects beyond the relatively limited timeframe of the trial[13].

This article details a joint consensus statement developed through the Forum on the Integration of Observational and Randomized Data (FIORD) meeting, held in Bethesda, Maryland on November 7-8, 2024. The meeting convened a multidisciplinary group of scientists and stakeholders affiliated with the U.S. Food and Drug Administration, the National Institutes of Health, the pharmaceutical industry, and academic research institutions, with expertise spanning clinical trials, causal inference, pharmacoepidemiology, and regulatory science. FIORD participants formed working groups to address a range of topics related to best practices for integrating RCTs and RWD to generate reliable evidence. The goal of this article is to help stakeholders understand how different objectives for RCT-RWD integration shape key design and analytic considerations. We outline practical issues that commonly arise, such as data relevance, curation, cross-source comparability, estimand choice, and sensitivity analysis. With the increasing availability of estimators that can be applied in RCT and RWD integration analyses, it can be tempting to conduct such analyses post hoc, treating data integration as a "savior" when trial power is low. We emphasize the importance of considering data integration strategies early in the study planning process, including, where feasible, the prospective collection of relevant RWD, pre-specification of the target population and estimand, and alignment of data capture across sources. This can ensure improved quality, transparency, and interpretability of the evidence generated. Data integration analyses already introduce substantial complexity. As a result, it can be useful to weigh the transparency, accessibility, and interpretability of candidate methods, alongside their assumptions and robustness, when planning and communicating an analysis, particularly in regulatory contexts**.** This article does not advocate replacing randomized evidence but rather provides guidance on principled integration when external data are relevant to a clearly defined causal estimand**.** Our goal is to build the case that principled integration of randomized and real-world evidence may enable more reliable treatment recommendations while preserving regulatory-grade evidentiary standards.

This article is organized as follows. Section 2 describes common goals of RCT-RWD integration and the types of scientific questions and study scenarios that external RWD may help address. Section 3 provides a review with illustrative examples of estimand definition in RCT-RWD integration settings. Section 4 discusses key challenges and design considerations for conducting such studies. Section 5 concludes.

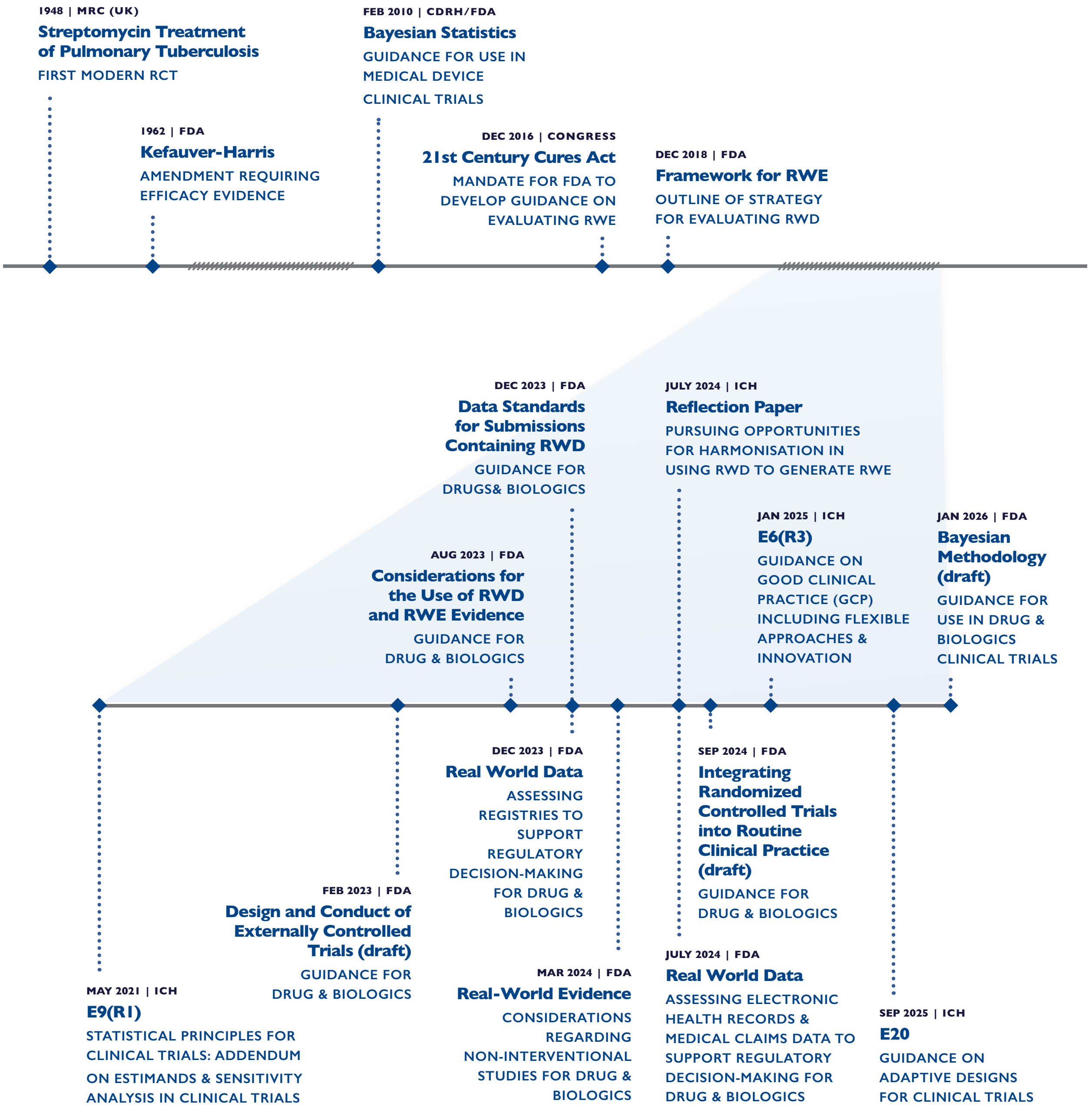

*Figure 1. Timeline of selected key milestones of regulatory guidance on real-world data (RWD) and real-world evidence (RWE), highlighting key historical, legislative, FDA, and ICH developments from 1948 to 2026.*

## 2 Objectives for integrating RCT and RWD to support regulatory decision-making

To promote transparency, ensure rigor in the evidence generation process, and support consistent evaluation of study quality, participants at the 2022 FIORD meeting identified the need for a unifying structure to help specify key elements of an RWE study's design and analysis plan[14]. One framework highlighted in this context is the *Causal Roadmap*, which can guide sponsors in designing studies and developing statistical analysis plans (SAPs), as well as help regulatory agencies and the public evaluate the quality of the evidence generated[14,15]. The Causal Roadmap provides a structured, iterative approach for formulating causal questions, defining target causal estimands, articulating identification assumptions, and selecting appropriate statistical estimators, with careful consideration of the data-generating processes and study design choices.

The first step of the Causal Roadmap requires researchers to understand the causal question of interest and goal of the study. Different objectives for RCT-RWD integration studies may inform which RWD source to use, choices of estimands, necessary identification assumptions, and estimators. To help illustrate the various goals of data integration studies, throughout the article, we use the DEVOTE RCT—a double-blind, treat-to-target, event-driven cardiovascular safety trial evaluating insulin degludec, an ultralong-acting, once-daily basal insulin, against insulin glargine—as a running example[16]. The DEVOTE trial demonstrated non-inferiority of degludec, meeting the prespecified hazard ratio margin of 1.3, indicating no more than a 30% increase in cardiovascular risk, relative to glargine.

Objectives for RCT-RWD integration can be grouped according to whether the goal is to enhance precision within the original causal estimand, modify the target population of inference, extend follow-up, or evaluate treatment performance under real-world conditions. Each objective may imply distinct choice of estimands along with their identification assumptions and design considerations.

1. **Improving statistical power:** Leveraging external RWD to augment the RCT could improve power for testing hypotheses on secondary endpoints, or sub-components of the primary composite endpoint[17]. For example, the DEVOTE RCT was adequately powered to test a non-inferiority hypothesis for its primary endpoint—a composite Major Adverse Cardiovascular Event (MACE) including cardiovascular death, nonfatal myocardial infarction, or nonfatal stroke. However, analyses of individual components are often underpowered because these events are rarer than the composite. In such cases, augmenting the RCT with external RWD, e.g., data from large-scale electronic health record (EHR) and insurance claims databases, may increase the number of observed events. Augmenting RCTs with external data may also improve the power of subgroup analyses[18]. For instance, in DEVOTE, the Asian subgroup was underrepresented, comprising only 10.2% of the trial sample[19]. Power gains must be weighed against potential bias introduced by systematic differences between RCT and RWD sources violating identification assumptions. Because these hybrid RCT-RWD designs retain randomized treatment and concurrent comparator arms, the RCT-only contrast remains protected by randomization and can serve as an internal benchmark when assessing or incorporating external RWD. This structure can limit the influence of bias arising from nonexchangeable external data by making the exchangeability assumption testable. In contrast, studies relying solely on a nonrandomized external comparator lack an internal randomized benchmark and generally require stronger, less verifiable exchangeability assumptions[3,20,21].
2. **Improving generalizability of trial findings in the target patient population:** *Generalizability* refers to the extent to which causal conclusions drawn from a study population can be validly applied to a broader population from which the study population was sampled (i.e., where the study population is a subset of the broader population)[10,11]. In this setting, the objective is to reweight or augment the RCT sample so that the distribution of baseline characteristics better reflects the target population. For example, the target population could be defined as all patients receiving routine care in the United States who would have met DEVOTE's eligibility criteria. Although these patients satisfy the same eligibility criteria as the trial participants, they may differ from enrolled participants in characteristics such as age, sex, race and ethnicity, duration of diabetes, kidney function, or history of cardiovascular disease because trial participation may be selective. Suppose, for instance, that chronic kidney disease is more common among eligible patients in routine practice than among DEVOTE participants. If kidney function modifies the comparative effect of insulin degludec versus insulin glargine, the average treatment effect in the trial-enrollment population may not equal the average effect in the broader trial-eligible population. RWD describing the eligible population could then be used to reweight the trial participants so that their baseline-characteristic distribution more closely represents that target population[8,22].

3. **Transporting trial findings to populations different from the trial:** RCT-RWD integration may also be used to study the causal effect of the experimental drug in some other populations of interest that are different from the trial population[11,23,24]. Broadly, generalization concerns extending inference from enrolled trial participants to a trial-eligible target population, whereas transportation concerns applying trial evidence to populations that may not be the same as the population from which trial participants were sampled, typically requiring stronger and often non-testable identification assumptions in practice[8]. In many cases, the population enrolled in an RCT is deliberately chosen to address a specific regulatory or scientific question but may differ in composition from the population of interest for subsequent clinical or policy decisions.

For example, the DEVOTE trial, like many cardiovascular safety trials, focused on individuals at high cardiovascular risk, reporting that 63% of participants had established cardiovascular disease. Yet decision-makers may also be interested in treatment effects among patients at lower cardiovascular risk who are more representative of routine practice. In such settings, integrating external RWD may potentially facilitate the estimation of treatment effects in such target populations that are different from the RCT, while making explicit the additional assumptions required for transporting trial findings beyond the original trial population (e.g., assumptions ruling out unmeasured effect modifiers after conditioning on measured baseline covariates)[23].

4. **Understanding long-term health outcomes:** RCTs typically span a relatively short period, making it difficult to study long-term clinical outcomes, especially for safety evaluation. To address this, researchers may link RCT participants to their future medical records in EHR databases (or match them with real-world patients sharing similar baseline characteristics), utilizing the post-trial RWD generated by the same group of trial patients. This enables the assessment of long-term drug effects and other extended outcomes beyond the trial's original follow-up period, including long-term safety monitoring of the drug. This is particularly valuable in settings where the event of interest is rare within the trial timeframe, such as the MACE example discussed under the first objective, or when addressing scientific questions about treatment effects over longer horizons, such as at 5 or 10 years, which would otherwise be infeasible within the constraints of the original trial. In some settings, longer-term comparative effectiveness questions may still be plausible, for example, when the RCT compares two active, clinically acceptable treatments rather than a placebo, and a substantial subset of patients continues their originally assigned therapy after trial completion. In the context of the DEVOTE trial, for example, linking participants to EHR or claims data after trial completion could potentially allow researchers to assess whether differences in cardiovascular outcomes between the investigational and comparator drug emerge or widen over time among patients who remain on their initial treatments, while also enabling long-term safety surveillance[25].

5. **Constructing an external comparator when an internal control is not feasible**: In some therapeutic areas, such as rare diseases, certain oncology indications, or settings where withholding treatment is not feasible, researchers often rely on single-arm trials that lack an internal control group[3,20]. In these cases, external controls constructed from RWD are often used to enable comparative effectiveness or safety analyses[26,27]. To improve comparability, investigators can attempt to emulate a target randomized trial as closely as possible by aligning key design elements including eligibility criteria, time zero, treatment assignment and outcome definitions between the internal treatment arm and the external RWD control cohort[3,28]. Importantly, selecting external controls based on post-treatment information, such as similar lengths of follow-up as the internal patients, is problematic and could induce selection bias. Unlike in the RCT-RWD integration setting where an internal randomized control arm is present (even if smaller in sample size compared to the internal treatment arm) and can be used to test for bias by comparing internal and external controls, single-arm designs offer no such internal check[29]. As a result, analyses that rely on external controls in single-arm trials are generally more challenging and would require stronger assumptions that are difficult to test using data.

These objectives are conceptually distinct but not mutually exclusive. A single RCT-RWD study may pursue multiple of these objectives. For example, integrating an RCT with a large observational cohort may both improve statistical precision and generalize the treatment-effect estimate to a trial-eligible target population[30]. Similarly, linking trial participants to longitudinal claims or electronic health record data may extend follow-up while increasing the number of observed events, thereby improving precision for rare or delayed outcomes[31].

# 3 Estimands for RCT-RWD integration

Once the scientific objective has been established and the target population has been defined as part of the research question, the next step in the Causal Roadmap is to formally specify a causal estimand that aligns with the objective and the underlying scientific question. In a general causal inference setting, this involves defining a Structural Causal Model (SCM)[32] that generates the observed data and specifying an intervention to define post-intervention distributions. In this section, we focus on examples of relevant causal estimands and strategies for defining them in the context of RCT-RWD integration. Different estimands may require different identification assumptions (i.e., assumptions on the underlying processes that generated the data, including assumptions on differences between these processes between data sources, needed to reliably recover the desired causal effect from the integrated data). Distinct causal estimands may also have nuanced differences in their interpretation, making them suitable for addressing different scientific or policy questions[33].

We begin by introducing notation for a point-treatment individual-level data structure in the context of RCT-RWD integration. Let the observed data be denoted by $O = (W, S, A, Y)$, where $W$ is a vector of baseline covariates, $A$ is a binary treatment indicator, and $Y$ is the outcome of interest. The variable $S$ indicates the data source where the patient comes from. Note that the objectives described in Section 2 encompass several distinct data-integration settings, and no single illustrative data structure is intended to represent all of them. To keep the following notation and estimand examples focused, we restrict attention to a population of trial-eligible patients in the external data. Thus, patients who would not have met DEVOTE's eligibility criteria are outside the target population of this particular example. Then, we can take $S \in \{0,1\}$, where $S = 1$ denotes participants in the RCT, and $S = 0$ denotes trial-eligible individuals who did not enroll in the RCT. For example, in our motivating case study of cardiovascular risk among diabetes drug users, $S = 1$ could correspond to participants in the DEVOTE trial, and $S = 0$ could represent individuals who met DEVOTE's eligibility criteria but did not participate in the trial. Here, we consider the general case where in the external data, there are both patients on the treatment and control. It covers the special case where we only have patients on the control arm in the $S = 0$-study. Suppose we have $n$ copies of $O$: $O_1, \cdots, O_n$ drawn i.i.d from the true probability distribution $P_0$ that lies in the nonparametric statistical model for patients $1, \ldots, n$.

The point-treatment structure is used here for illustration purposes and should not be understood as the sole data structure that arise in RCT-RWD integration. In chronic treatment settings, such as basal insulin therapy, $A$ may represent initial treatment assignment or initiation, whereas the full clinically relevant exposure may involve a sequence of treatment decisions over time, including dose titration, adherence, discontinuation, switching, and use of concomitant therapies. In such settings, the data structure may be modified to reflect the underlying data-generating experiment, and the causal estimand may need to be defined in terms of longitudinal treatment regimens rather than a single time-point intervention. For example, one might compare a strategy of initiating insulin degludec and continuing or titrating treatment according to prespecified clinical rules versus an analogous strategy for insulin glargine over a defined follow-up period. Formalizing such questions would require a longitudinal data structure, for example, $O = (W, S, L_0, A_0, C_1, Y_1, L_1, A_1, C_2, Y_2, \ldots, C_K, Y_K)$, where $L_t$ includes time-varying covariates, $A_t$ denotes treatment at time $t$, $Y_{t+1}$ is an event indicator during interval $(t, t+1]$, e.g., MACE, and $C_{t+1}$ denotes censoring or loss-to-follow-up during interval $(t, t+1]$.

## 3.1 Strategies for defining potential outcomes

We assume a SCM defined by: $W = f_W(U_W)$, $S = f_S(W, U_S)$, $A = f_A(W, S, U_A)$, and $Y = f_Y(W, S, A, U_Y)$, for unspecified functions $f_W$, $f_S$, $f_A$, $f_Y$, where $U = (U_W, U_S, U_A, U_Y)$ is a vector of errors. In other words, we assume that our observed data were generated in a process in which baseline characteristics may have affected both treatment assignment and outcome, treatment may have affected outcome (the effect of interest), and both processes may have differed between studies; further, baseline characteristics may have affected RCT participation. The ordering of baseline covariates $W$ affecting trial participation $S$ reflects our definition of $W$ as characteristics that exist before trial participation is determined. To illustrate, consider a source population restricted to patients who meet the DEVOTE eligibility criteria. Here, $W$ may include age, sex, duration of diabetes, cardiovascular disease history, and kidney function; $S$ indicates enrollment in DEVOTE; $A$ denotes treatment with insulin degludec or insulin glargine; and $Y$ denotes a subsequent MACE outcome. Baseline characteristics may influence trial participation—for example, cardiovascular risk, geographic location, or access to trial site may affect enrollment—and therefore $W$ precedes and may cause $S$. This ordering does not imply that the observed distribution of $W$ is the same across sources. Selection into the trial can produce different distributions of $W$ conditional on $S$. It also assumes that $W$ represents underlying pre-participation characteristics rather than measurements themselves being caused by the data source.

We can define potential outcomes through hypothetical interventions on the $A$ node. Specifically, we define potential outcomes by replacing $A$ with $A = a$ for $a \in \{0,1\}$. This yields $Y_1 = f_Y(W, S, A = 1, U_Y)$ and $Y_0 = f_Y(W, S, A = 0, U_Y)$, the outcomes that would have been observed for a randomly sampled individual had they, possibly contrary to fact, been assigned to treatment or control, respectively. One must define an SCM that accurately represents how the data are (or might be) generated, including data from multiple sources in data integration problems. In the context of RCT-RWD data integration, interventions may be specified not only on the treatment and censoring nodes but also on other relevant nodes, such as those representing different studies/data sources. For example, one could additionally intervene on the study indicator $S$ node and define potential outcomes $Y_{11} = f_Y(W, S = 1, A = 1, U_Y)$ and $Y_{10} = f_Y(W, S = 1, A = 0, U_Y)$. These correspond to the potential outcomes under jointly enforcing assignment to the RCT setting and to treatment or control, respectively. Importantly, allowing the potential outcomes to depend on the study indicator $S$ captures the possibility that trial participation itself may influence outcomes through protocol adherence, monitoring intensity, or other contextual effects.

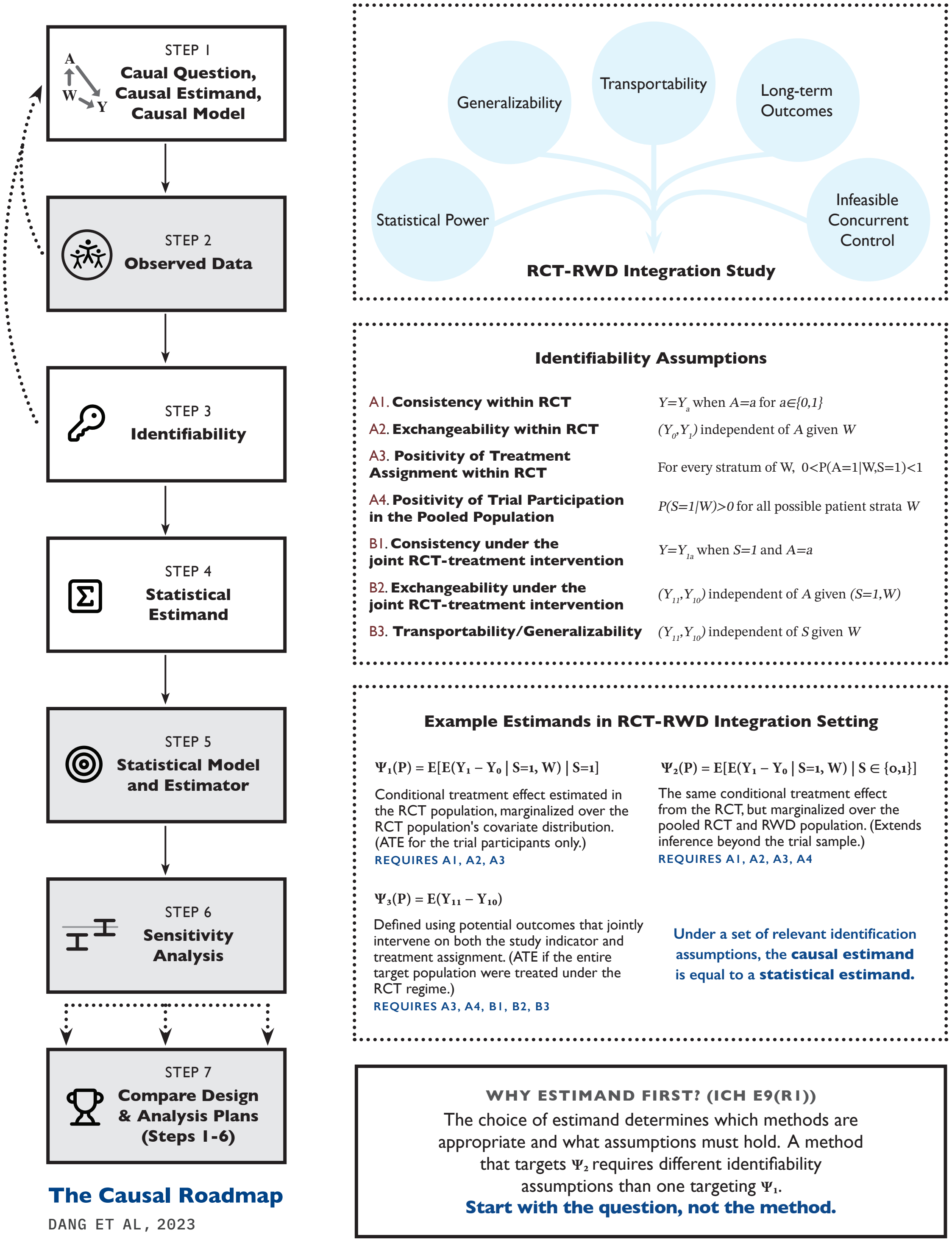

*Figure 2. The Causal Roadmap, objectives of RCT-RWD integration studies, causal estimands and their identification assumptions.*

## 3.2 Example estimands and their identification assumptions in RCT-RWD integration setting

A causal estimand is the specific, target causal effect one would like to estimate in a study; it is typically stated as a contrast between population-level summaries of the potential outcome under different treatment levels. For example, for a given population, the average treatment effect is defined as $E(Y_1) - E(Y_0)$, i.e., the difference in expected potential outcome under treatment vs. control. Identification assumptions allow the causal estimand to be re-expressed as a statistical estimand, which is a function of the observed data distribution and thus estimable using available data.

In RCT-RWD integration settings, one possible causal estimand is the average treatment effect within the original trial population, which can be formulated as $\Psi_1(P) = E_P[E_P(Y_1 - Y_0|S = 1, W)|S = 1]$; here, the conditional treatment effect is taken on the RCT population and then marginalized over the same RCT population. The following three assumptions are sufficient to nonparametrically identify the causal estimand $\Psi_1$:

A1. **(Consistency)** $Y = Y_a$ when $A = a$ for $a \in \{0,1\}$ in the RCT.

A2. **(Exchangeability within the RCT).** In the RCT, the potential outcomes $(Y_0, Y_1)$ are independent of the treatment assignment $A$ given baseline covariates $W$, i.e., $(Y_0, Y_1) \perp A|W, S = 1$.

A3. **(Positivity of treatment within the RCT)** For every stratum of observed $w$, $0 < P(A = 1|w, S = 1) < 1$. This ensures that every individual in the trial has a nonzero probability of receiving either treatment/control.

Note that assumptions A2 and A3 are supported by randomized assignment in the RCT, while A1 requires the treatment intervention to be sufficiently well defined. As an example, if the researcher wishes to make no additional assumptions, beyond the standard assumptions above commonly assumed to hold due to the design of an RCT, but still wishes to use the external RWD to improve power, then $\Psi_1$ may be considered.

If the researcher would also like to extend inference to the pooled RCT and RWD population, the estimand could instead be $\Psi_2(P) = E_P[E_P(Y_1 - Y_0|S = 1, W)|S \in \{0,1\}]$, which marginalizes over the pooled population. Importantly, in addition to the above mentioned three identification assumptions, estimand $\Psi_2(P)$ also requires the following support assumption:

A4. **(Positivity of trial participation)** $P(S = 1|w) > 0$ for all observed patient strata $w$ in the pooled population.

Note that the identification of estimand $\Psi_1$ does not require any external data; while the identification of $\Psi_2$ does not require treatment or outcome data from the external patients, as baseline covariates from the target population are sufficient to define the covariate distribution over which the RCT conditional treatment effect is averaged. Nevertheless, when external RWD include treatment and outcome information of adequate relevance and reliability, these data may improve finite-sample efficiency by, for example, helping estimate nuisance functions or conditional treatment effects better. For instance, one may include both RCT-only and pooled RCT-RWD outcome regression learners in a discrete Super Learner library and use cross-validated predictive performance evaluated on the RCT sample to determine whether external outcome information improves the outcome regression fit[34].

There are subtle differences between the two estimands $\Psi_1$ and $\Psi_2$, and they may not be equal in practice. Let $\tau_P(w) = E_P(Y|S = 1, A = 1, W = w) - E_P(Y|S = 1, A = 0, W = w)$ be the conditional treatment effect in the RCT. Under the corresponding identification assumptions, the

two estimands are equal to the statistical estimands $\Psi_1(P_0) = E_0[\tau_0(W)|S = 1]$ and $\Psi_2(P_0) = E_0[\tau_0(W)]$, respectively, where we use $E_0$ and $\tau_0$ to suppress notation $E_{P_0}$ and $\tau_{P_0}$, respectively. The two estimands coincide when the trial and target populations have the same distribution of effect modifiers, when $\tau_0(W)$ is constant, or when differences across effect-modifier strata cancel exactly. More generally, they differ when treatment effects vary across $W$ and the distribution of those effect modifiers differs between the trial and target populations. We provide three examples to help readers better understand and interpret the difference between the two estimands.

*Example 1: One effect modifier.* Suppose the risk difference comparing degludec with glargine is -1 percentage point among patients with established CVD and -3 percentage points among patients without established CVD. If 63% of DEVOTE participants have established CVD but only 40% of the target population does, then $\psi_1 = 0.63(-1) + 0.37(-3) = -1.74$, where as $\psi_2 = 0.40(-1) + 0.60(-3) = -2.20$. Thus, $\psi_2$ indicates a larger average benefit because the target population contains more patients in the stratum with the larger treatment effect.

*Example 2: Differences reinforce each other.* Suppose established CVD and impaired renal function both attenuate the benefit of degludec. If the target population has lower prevalences of both characteristics than DEVOTE, both distributional differences would tend to make $\psi_2$ more negative than $\psi_1$.

*Example 3: Differences offset each other.* Suppose established CVD and older age both attenuate treatment benefit. If the target population has fewer patients with established CVD but is older than DEVOTE, the first difference would tend to make $\psi_2$ more negative than $\psi_1$, whereas the second would tend to make it less negative. The net difference would depend on the magnitude of effect modification and the joint distribution of age and CVD.

Alternatively, using potential outcomes $Y_{11}$ and $Y_{10}$ (i.e., the outcomes under a hypothetical intervention to enroll everyone in the pooled target population into the RCT, and then assign treatment or control), a possible causal estimand could be $\Psi_3(P) = E_P(Y_{11} - Y_{10})$. The following assumptions, along with assumptions A3 and A4, are sufficient for nonparametric identification of $\Psi_3$:

B1. **(Consistency)** $Y = Y_{1a}$ when $S = 1$ and $A = a$.

B2. **(Exchangeability within the RCT)** $(Y_{11}, Y_{10})$ independent of $A$ given $(S = 1, W)$.

B3. **(Transportability/generalizability across studies)** $(Y_{11}, Y_{10})$ independent of $S$ given $W$.

Note that Assumption B3 suggests that conditioning on the measured baseline covariates, trial participation carries no additional information about how a patient would respond under the RCT setting. That is, the measured baseline covariates capture all differences between the trial participants and non-participants that matter for the potential outcomes under the RCT regime, including effect modification and prognosis under each arm.

As an example, suppose the target population is DEVOTE-eligible patients with type 2 diabetes in the US routine care. The estimand $\Psi_1$ averages the risk difference over the DEVOTE trial population. In contrast, $\Psi_3$ can be interpreted as the average risk difference that would be observed if the broader DEVOTE-eligible routine-care population were enrolled in a DEVOTE-like trial setting and assigned degludec versus glargine under the trial protocol. This estimand could differ from $\Psi_1$ if the routine-care population contains more patients with effect-modifying characteristics, such as chronic kidney disease, older age, or different cardiovascular risk profiles. It could also differ from an effect estimated directly under routine-care conditions if treatment adherence, dosing, monitoring intensity, or outcome ascertainment differ between routine care and the trial setting. Thus, $\Psi_3$ is useful when the scientific question concerns the effect expected under the RCT regime in the broader target population, rather than the effect of treatment as implemented in routine care.

To summarize, the estimands $\Psi_1$ and $\Psi_2$ leverage randomization within the RCT to establish nonparametric identification and differ primarily in the baseline covariate distribution over which the conditional treatment effect in the RCT, i.e., $\tau(W)$, is averaged (e.g., trial-enrolled population only versus a broader trial-eligible population). Importantly, without additional assumptions about trial participation, these estimands retain a clear causal interpretation within the trial but do not, in general, correspond to the causal effect of intervening to treat the broader real-world population under the RCT setting. Interestingly, under the identification assumptions discussed above for $\Psi_3$, we can express it as $\Psi_3(P_0) = E_0[\tau_0(W)]$, which is equal to $\Psi_2(P_0)$. In other words, if one is willing to assume additionally assumption B3 (transportability/generalizability across studies), then the causal interpretation of $\Psi_2$ is enriched and can be interpreted as the average effect of assigning individuals in the target population to treatment versus control under the RCT regime. This assumption is substantively stronger, requiring that all factors that jointly influence trial participation and potential outcomes under the trial setting are captured, but it enables a sharper policy-relevant statement about the effect that would be expected if the target population were treated according to the trial protocol.

# 4 Designing RCT-RWD integration studies: challenges and considerations

When integrating external data with RCTs, it is important to ensure that the external RWD are fit-for-purpose for the intended scientific and regulatory objective. Consistent with FDA's guidance *Real-World Data: Assessing Electronic Health Records and Medical Claims Data to Support Regulatory Decision-Making for Drug and Biological Products*, we organize key considerations into data relevance and data reliability. In RCT-RWD integration settings, these considerations also include how well core design elements and measurements can be aligned across the trial and routine-care data sources[3,35,36].

1. **Data relevance (fit-for-purpose):** Relevance concerns whether the external data source is appropriate for the study question and estimand, i.e., whether it contains the necessary information on the right patients, treatments, outcomes, and baseline covariates, to support the intended analysis. Practical considerations include how well the external cohort aligns with the target population defined by the estimand (e.g., whether trial eligibility criteria can be emulated in RWD using available variables) and timing of data collection (e.g., whether the data are concurrent with the trial or drawn from earlier/later periods). When external data are not collected concurrently, studies sometimes draw on historical controls or incorporate post-market data, which can raise additional compatibility questions, such as whether changes in standard-of-care or outcome assessments may affect interpretation. More broadly, the external data's relevance to the target population matters. For example, when the goal is to enhance generalizability, the combined RCT and RWD sample should reflect the population to which conclusions are intended to apply, i.e., the target population defined by the estimand.
2. **Data reliability (quality and integrity):** Reliability concerns whether the data are sufficiently accurate, complete, and traceable, both in the source data and after curation and construction of the analytic dataset, to support the proposed design and analysis. This includes evaluating whether the external data capture key variables with sufficient completeness and quality to support the intended assumptions, including patterns of missingness, measurement error, and potential biases in data recording. Besides standard data quality checks, RCT-RWD integration settings raise additional concerns about whether variables are measured in comparable ways across sources. For example, trials may use adjudicated endpoints, whereas real-world databases may rely on routine documentation and coding, which can lead to differences in outcome ascertainment (including potential underreporting) and timing precision. Differences in adherence and treatment switching between trial protocols and routine care can also create mismatches in exposure definitions. Specifically, a coarsened treatment node $A$ may not define a unique intervention if outcomes depend on these different versions of treatment. Patients in RCTs may receive more intensive monitoring and protocol-directed care than patients in routine practice. For example, DEVOTE used treat-to-target insulin titration and adjudicated MACE outcomes, whereas RWD may reflect less standardized titration and identify MACE using diagnosis or procedure codes[16,36]. In addition, inclusion and continued observation in RWD may depend on insurance coverage or encounters within a healthcare system. These differences may introduce source-specific measurement error or informative selection, even in the absence of unmeasured treatment-outcome confounding.

## 4.1 Sources of bias and assessment considerations

In Table 1, we summarize common sources of misalignment between an RCT and external RWD, along with guiding questions and practical checks that can be considered when designing a data integration study and reported to help relevant parties assess potential sources of bias[3,28,35–42]. For example, calendar-time mismatch can introduce bias through nonconcurrency (e.g., changes over time in clinical practice, especially if the control arm is standard-of-care, and outcome assessment)[43]. Restricting to strictly concurrent RWD may reduce the bias, but it can also leave too few external patients to meaningfully improve precision (i.e., variance reduction). In practice, expanding the calendar window often increases sample size and event counts, but may amplify bias.

| Potential RCT & RWD Misalignment | Primary Consideration | Assessment Considerations |
|---|---|---|
| Target population / eligibility criteria differences | Relevance | - Can we emulate trial eligibility in RWD with available variables?<br>- Are there any criteria that are not measured, missing, or measured differently in the RWD (e.g., lab measurements, patient characteristics)?<br>- Compare post-eligibility baseline covariate distributions (e.g., standardized mean differences, histograms) and document any differences. |
| Calendar-time mismatch (historical vs. concurrent controls) | Relevance | - Plot (or compute summary statistics of) key baseline covariates, outcome incidence by calendar time in RWD, check whether trial periods have good overlap.<br>- Did guidelines, coding systems, or treatment options change across time periods? Consider sensitivity analysis if relevant.<br>- Consider restricting RWD period to concurrent windows. |
| Care setting differences (trial sites vs routine care) | Relevance (and may inform reliability) | - Compare proxies for care intensity (e.g., follow-up frequency, lab frequency, hospitalization coding density).<br>- Does monitoring intensity differ enough to change outcome detection? |
| Baseline covariate definition differences | Reliability | - Harmonize code lists, units, time windows (e.g., “most recent lab within 90 days”).<br>- Assess missingness and potential measurement error.<br>- Document harmonization decisions and traceability. |
| Baseline covariate distribution differences | Relevance | - Check distributions and outliers in RWD, assess comparability with the RCT.<br>- Is there sufficient overlap in covariate distributions between RCT and RWD?<br>- Matching based on baseline covariates or summary measures, e.g., probability of trial-enrollment given baseline covariates. |
| Treatment definition differences | Reliability | - In RWD, check treatment initiation rules, grace periods, switching, and adherence.<br>- Are there region-specific care guidelines?<br>- Document any differences in treatment definition between RCT and RWD.<br>- Consider sensitivity analysis using negative control treatments. |
| Outcome definition differences (adjudicated endpoints vs claims/EHR codes) | Reliability | - Compare outcome rates, could RWD source systematically under-detect events?<br>- Compare follow-up visit rates and coding density.<br>- Assess date precision: are there any differences in endpoint timing granularity between RCT and RWD?<br>- Check claims runout or EHR completeness near the end of observation. Are recent events undercounted due to the possibility of a lag? Restrict to periods with complete data.<br>- Consider sensitivity analysis using negative control outcomes. |

**Table 1.** Potential RCT & RWD misalignment with corresponding list of assessment considerations.

As another example, when target populations differ, investigators often use matching to improve comparability between the RCT and the external RWD cohort. If matching is used to define the population over which effects are averaged, then the target estimand may shift to the matched population and should be stated explicitly. In contrast, matching can also be used only as an auxiliary design step to select external RWD patients who are similar to RCT participants for the purpose of estimating nuisance functions, such as outcome regressions or conditional treatment effects. In that case, the target estimand can remain the RCT-population effect (e.g., estimand $\Psi_1$ in our example in Section 3). In practice, even after matching on observed baseline covariates (or on their summary statistics), external patients who otherwise meet trial eligibility criteria may still differ from the trial sample in, e.g., variables that reflect health seeking behavior, care settings, or other unobserved pre-treatment characteristics. Some of these differences may be hypothesized to be weak confounders or clinically less relevant. Clearly documenting these baseline differences and the rationale for any inclusion criteria is important.

A related design consideration is that the selection of an external RWD cohort should be based only on information available at baseline. Conditioning on post-baseline information, such as future follow-up duration, treatment changes, or intermediate outcomes, may introduce selection bias related to survival, continued enrollment, or engagement with care. For example, although it may be tempting to restrict the external RWD cohort to patients with at least one year of follow-up to “match” the trial’s follow-up, such a restriction implicitly conditions on being event-free and observable for that period and can distort the risk set and outcome process, i.e., it creates a form of immortal time or selection bias.

Differences in how treatment is defined and implemented are another common challenge in RCT-RWD integration. For example, RCT protocols may restrict treatment switching and closely monitor adherence, whereas in routine care such controls are typically not enforceable, and patients may discontinue, switch, or add additional therapies over time. As a result, the definition of treatment in RWD is often operational, e.g., classifying patients as treated if they initiate a drug within a prespecified window, yet this may correspond to a more heterogeneous set of exposures (initiation, partial adherence, switching patterns) than the protocol-defined intervention in the RCT or more susceptible to exposure misclassification. In turn, the relationship between the RWD exposure definition and the trial intervention can bear on the plausibility of assumptions commonly used for causal identification, including the consistency assumption, which states that there is only one version of treatment[44]. From an analytic perspective, different data integration estimators may handle this mismatch differently: some approaches treat discrepancies in treatment implementation as part of the overall bias to be estimated, while others rely more directly on alignment of the intervention and adherence patterns across data sources and may therefore be more sensitive to such differences. These issues could be partially addressed by representing the observed data using a longitudinal data structure with repeated measurements of time-varying covariates, treatments, censoring, and outcomes. The causal estimand could then contrast potential outcomes under sustained or dynamic treatment regimens defined by joint interventions on multiple treatment nodes over time. Although general methods for longitudinal causal inference are well developed, methods specifically tailored to integrating RCT and RWD for longitudinal treatment-regimen estimands remain comparatively limited. Moreover, a longitudinal representation does not itself resolve cross-source differences or violations of sequential exchangeability and positivity.

An important consideration is the role of sensitivity analyses in RCT-RWD integration studies. Sensitivity analyses are most informative when they complement a carefully designed primary analysis. Even with a strong primary analysis, the plausibility of certain identification assumptions may remain unknown. Sensitivity analyses can then be used to assess how conclusions might change under possible and relevant deviations from these assumptions, and to communicate the robustness of the primary analysis results to stakeholders. At the same time, one may interpret certain “fixes” placed only in sensitivity analyses as signaling that the primary analysis did not fully engage with an important source of bias that could have been addressed directly. In that sense, there is value in reserving sensitivity analyses for uncertainty that cannot be eliminated through study design or analysis methods alone. For example, when event or censoring models are specified using a Cox proportional hazards model in the primary analysis, a sensitivity analysis that relaxes the proportional hazards or functional form assumptions (e.g., via more flexible hazard models) may reveal sensitivity to model misspecification and even yield conflicting results. However, such flexibility is often more appropriately incorporated directly into the primary analysis, for instance, by using targeted maximum likelihood estimation with Super Learner for flexible nuisance estimation, rather than deferred to a sensitivity check[45]. If the goal of RCT-RWD integration is to improve statistical power (for secondary outcomes), one may also consider reporting 1) the RCT-only estimate, 2) a naive pooled analysis that combines RCT and RWD without applying any data integration estimators that account for bias, and 3) one or more data integration analyses that vary key design choices (e.g., calendar window, matching specifications, truncation thresholds for any inverse weights). Presenting these side-by-side can clarify how much the conclusions are driven by the data integration versus the trial evidence alone and can help diagnose whether gains in precision come at the cost of noticeable shifts in effect estimates[46,47].

Negative controls may also arise either as part of an estimation strategy or as part of a sensitivity analysis. For example, some frameworks (e.g., proximal causal inference) incorporate negative controls directly into identification and estimation[48], and some RCT-RWD integration estimators explicitly leverage negative control information[49]. Another example is the G-value approach[50], which frames sensitivity analysis in terms of a causal gap, i.e., differences between the causal estimand of interest and the statistical estimand identified under a given set of assumptions. The corresponding G-value summarizes the minimum causal gap that would be required to overturn a study conclusion (e.g., to move a statistically significant result to include the null)[14,51,52].

Overall, the checklist provided in Table 1 is intended to support transparent documentation, communication, and interpretation of the most plausible sources of bias in RCT-RWD integration. The aim is to make one aware of key design and data-quality considerations, encourage clear reporting of how each issue was assessed and addressed, and provide context for interpreting results. The degree to which any misalignment is consequential can depend on the data analysis method, since different methods vary in the assumptions they make.

## 4.2 Estimators for RCT-RWD data integration

A wide range of data integration estimators has been developed for RCT-RWD integration. An important consideration is that different estimators not only rely on different assumptions but may target different estimands. As a result, direct comparisons across methods can be challenging when the underlying estimands differ. With this caveat in mind, we provide a brief overview of several classes of existing approaches. Detailed studies of these methods on their assumptions, theoretical properties, and empirical performance are beyond the scope of this article. We refer readers to Shi et al.[8], Colnet et al.[9], Lin et al.[53] for more detailed technical reviews of statistical methods for analyzing RCT-RWD integration studies, and Degtiar and Rose[54] for a review focused on generalization and transportation methods.

One class consists of test-then-pool methods[55]. These approaches first conduct a hypothesis test comparing outcomes between the external arm and the RCT's internal arm. If no statistically significant difference is detected, the external and internal controls are pooled for analysis. Otherwise, inference is based solely on the RCT data. A practical limitation is that failing to detect a difference does not guarantee exchangeability, particularly when the RCT control arm is small. Several extensions and refinements of this idea have been proposed[21,49,56,57].

A second class comprises Bayesian and other dynamic borrowing methods. Power prior approaches incorporate historical data through a discount factor that down-weights the likelihood contribution of external patients[58] and its variants[59]. Commensurate priors assume that treatment effects in the trial and external data arise from a hierarchical model, effectively shrinking external information toward the RCT estimate when conflicts arise[60]. The meta-analytic-predictive prior constructs a prior for the trial effect by synthesizing historical data within a meta-analytic framework[61].

A third class includes bias-correction approaches. For instance, in adaptive TMLE[62], one first estimates the pooled estimand, then explicitly defines the corresponding bias estimand, estimates this bias, and finally applies a correction. The amount of efficiency gains from incorporating external data is driven by both the magnitude of the bias and complexity of the conditional effect of the study indicator on the outcome.

More recently, methods have been proposed that adaptively select external data rather than borrowing from all available sources. Conformal selective borrowing uses conformal prediction techniques to identify external patients whose covariate profiles suggest a high degree of exchangeability with the trial cohort. Selective borrowing procedures can improve robustness but require transparent reporting of inclusion criteria and assessment of how selection may alter the target population. For example, Zhu et al.[63] and Liu et al.[64] propose computing nonparametric conformity scores for each external patient and include only those with high similarity. These are effectively individualized test-the-pool procedures that operate at the patient level rather than at the study level.

# 5 Discussion

Interest in integrating RCTs with RWD continues to grow as stakeholders seek evidence that is both internally valid and relevant to clinical practice. Across the objectives discussed in this paper: improving power, enhancing generalizability, transporting trial findings, extending follow-up, and constructing external controls when concurrent controls are infeasible, the credibility of RCT-RWD integration studies is closely tied to how well the scientific objective, estimand, identification assumptions, data sources, estimators, and sensitivity analyses align. Framing these studies through the Causal Roadmap helps make these dependencies explicit and provides a common language for communicating how design and analytic choices map to the causal question of interest and the population to which the conclusions are intended to apply.

We discussed example estimands in the context of RCT-RWD integration and highlighted that even closely related target estimands can have meaningfully different interpretations. For example, one may consider averaging the conditional treatment effect in the trial over the RCT covariate distribution versus over a pooled covariate distribution. These estimands coincide only under special circumstances, most notably, when baseline covariates do not modify the treatment effect or when the covariate distributions are the same across sources, and otherwise represent different summaries of potentially heterogeneous treatment effects. As discussed in Section 3.2, the interpretation of certain estimands may also be enriched under additional assumptions about trial participation (e.g., that measured baseline covariates capture all effect modifiers relevant to differences between participants and non-participants). For this reason, it is helpful to state the estimand and its identification assumptions in terms of 1) the treatment effect being targeted and 2) the population distribution over which the contrast is averaged. This is particularly important in RCT-RWD integration studies, where multiple estimands may appear equally natural but address different scientific questions.

We note several limitations of this consensus-based overview. First, the considerations summarized in Table 1 are not exhaustive, and the relevance of any given item can vary by study objective, therapeutic area, endpoint type, and data provenance. Second, this paper does not recommend nor attempt to evaluate or compare specific estimators for RCT-RWD integration. Ongoing methodological work continues to explore ways to borrow information adaptively while guarding against bias when misalignments between data sources are suspected to occur, including approaches that explicitly quantify and adjust for residual bias, and approaches that restrict borrowing to subsets of external data that appear most comparable to the trial. Progress in method development may also be supported by shared benchmarking datasets, beyond small-scale isolated simulations, that could help standardize evaluation of the estimators.

To summarize, there is growing opportunity for prospective hybrid designs, in which relevant RWD are identified, curated, and quality-checked early, enabling clearer alignment of key data and design elements. Examples may include clinical trials which pre-specify in the protocol the RWD source such as an external registry, concurrent EHR curation, or post-trial linkage. RCT-RWD integration has the potential to expand the evidentiary toolkit available to regulators, clinicians, researchers, and patients, but its value depends on clear articulation of the study objective, careful specification of estimands, transparent justification and assessment of identification assumptions, and explicit documentation of data source alignment. Clear articulation of estimands and assumptions should support more consistent regulatory evaluation of RCT-RWD integration studies across therapeutic areas. The checklist and examples presented in this article are intended to support that transparency and to facilitate more consistent communication about what RCT-RWD integration studies can, and cannot, be interpreted in each setting.

## Acknowledgement

The authors express their sincere gratitude to Aimee Harrison and Alex Steele for their assistance in designing the figures for this manuscript.